\documentclass[usenatbib]{mn2e}
\usepackage{psfig}
\usepackage{times}
\usepackage{amssymb}
\usepackage{graphicx}
\usepackage{graphics}

\def\apj{ApJ} \def\apjl{ApJL} \def\mnras{MNRAS} \def\pasp{PASP}
  \def\araa{ARAA} \def\aap{A\&AP}
\def\aj{AJ} \def\apjs{APJS}  \def\nat{Nature}

\def\gs{\mathrel{\raise0.35ex\hbox{$\scriptstyle >$}\kern-0.6em
\lower0.40ex\hbox{{$\scriptstyle \sim$}}}}
\def\ls{\mathrel{\raise0.35ex\hbox{$\scriptstyle <$}\kern-0.6em
\lower0.40ex\hbox{{$\scriptstyle \sim$}}}}

\def\kms{\,\hbox{km}\,\hbox{s}^{-1}}

\def\Wm2{\,\hbox{W}\,\hbox{m}^{-2}}
\def\gsim{\mathrel{\raise0.35ex\hbox{$\scriptstyle >$}\kern-0.6em\lower0.40ex\hbox{{$\scriptstyle \sim$}}}}
\def\lsim{\mathrel{\raise0.35ex\hbox{$\scriptstyle <$}\kern-0.6em\lower0.40ex\hbox{{$\scriptstyle \sim$}}}}
\def\ltsima{$\; \buildrel < \over \sim \;$}
\def\simlt{\lower.5ex\hbox{\ltsima}}
\def\gtsima{$\; \buildrel > \over \sim \;$}
\def\simgt{\lower.5ex\hbox{\gtsima}}

\begin{document}

\title[ALMA insight on star formation in a distant protocluster]{ALMA
  observations of a $z\approx$~3.1 Protocluster: Star Formation from
  Active Galactic Nuclei and Lyman-Alpha Blobs in an Overdense
  Environment}

\author[D.~M.~Alexander et al.]
{ \parbox[h]{\textwidth}{ 
D.\ M.\ Alexander,$^{\! 1\, *}$
J.\ M.\ Simpson,$^{\! 1}$
C.\ M.\ Harrison,$^{\! 1}$
J.\ R.\ Mullaney,$^{\! 2}$
I. Smail,$^{\! 1}$
J.\ E.\ Geach,$^{\! 3}$
R.\ C.\ Hickox,$^{\! 4}$
N.\ K.\ Hine,$^{\! 3}$
A.\ Karim,$^{\! 5}$
M.\ Kubo,$^{\! 6,7}$
B.\ D.\ Lehmer,$^{\! 8}$
Y.\ Matsuda,$^{\! 9,10}$
D.\ J.\ Rosario,$^{\! 1}$
F.\ Stanley,$^{\! 1}$
A.\ M.\ Swinbank,$^{\! 1}$
H.\ Umehata,$^{\! 11,12}$
and T.\ Yamada$^{\! 13}$
}
\vspace*{6pt} \\
$^1$Centre for Extragalactic Astronomy, Department of Physics, Durham University, South Road, Durham, DH1 3LE, UK \\
$^2$Department of Physics and Astronomy, The University of Sheffield, Hounsfield Road, Sheffield, S3 7RH, UK\\
$^3$Centre for Astrophysics Research, Science and Technology Research Institute, University of Hertfordshire, Hatfield AL10 9AB, UK\\
$^4$Department of Physics and Astronomy, Dartmouth College, 6127 Wilder Laboratory, Hanover, NH 03755, USA\\
$^5$Argelander-Institut f\"ur Astronomie, Universit\"at Bonn, Auf dem H\"ugel 71, D-53121 Bonn, Germany\\
$^6$Institute for Cosmic Ray Research, University of Tokyo, 5-1-5 Kashiwa-no-Ha, Kashiwa City, Chiba 277-8582, Japan\\
$^7$Astronomical Institute, Tohoku University, 6-3 Aoba, Aramaki, Aoba-ku, Sendai, Miyagi 980-8578, Japan\\
$^8$University of Arkansas, 226 Physics Building, Fayetteville, AR 72701, USA\\
$^9$National Astronomical Observatory of Japan, 2-21-1 Osawa, Mitaka, Tokyo 181-8588, Japan\\
$^{10}$Graduate University for Advanced Studies (SOKENDAI), Osawa 2-21-1, Mitaka, Tokyo 181-0015, Japan\\
$^{11}$European Southern Observatory, Karl-Schwarzschild-Str. 2, D-85748 Garching, Germany\\
$^{12}$Institute of Astronomy, School of Science, The University of Tokyo, 2-21-1 Osawa, Mitaka, Tokyo 181-0015, Japan\\
$^{13}$Astronomical Institute, Tohoku University, 6-3 Aoba, Aramaki, Aoba-ku, Sendai, Miyagi 980-8578, Japan\\
$^*$Email: d.m.alexander@durham.ac.uk \\
}
\maketitle
\begin{abstract}

We exploit ALMA 870~$\mu$m observations to measure the star-formation
rates (SFRs) of eight X-ray detected Active Galactic Nuclei (AGNs) in
a $z\approx$~3.1 protocluster, four of which reside in extended
Ly~$\alpha$ haloes (often termed Ly~$\alpha$ blobs: LABs). Three of
the AGNs are detected by ALMA and have implied SFRs of
$\approx$~220--410~$M_{\odot}$~yr$^{-1}$; the non detection of the
other five AGNs places SFR upper limits of
$\simlt$~210~$M_{\odot}$~yr$^{-1}$. The mean SFR of the protocluster
AGNs ($\approx$~110--210~$M_{\odot}$~yr$^{-1}$) is consistent (within
a factor of $\approx$~0.7--2.3) with that found for co-eval AGNs in
the field, implying that the galaxy growth is not significantly
accelerated in these systems. However, when also considering ALMA data
from the literature, we find evidence for elevated mean SFRs (up-to a
factor of $\approx$~5.9 over the field) for AGNs at the protocluster
core, indicating that galaxy growth is significantly accelerated in
the central regions of the protocluster. We also show that all of the
four protocluster LABs are associated with an ALMA counterpart within
the extent of their Ly~$\alpha$ emission. The SFRs of the ALMA sources
within the LABs ($\approx$~150--410~$M_{\odot}$~yr$^{-1}$) are
consistent with those expected for co-eval massive star-forming
galaxies in the field. Furthermore, the two giant LABs (with physical
extents of $\simgt$~100~kpc) do not host more luminous star formation
than the smaller LABs, despite being an order of magnitude brighter in
Ly~$\alpha$ emission. We use these results to discuss star formation
as the power source of LABs.

\end{abstract}


\section{Introduction}

A key goal of observational cosmology is to understand how galaxies
and massive black holes (BHs) grow as a function of
environment. Models of large-scale structure formation predict that
galaxy and BH growth in distant high-density regions will be
accelerated in comparison to the growth of systems in typical regions
of the distant Universe (i.e.,\ the field; e.g.,\ Kauffmann
et~al. 1997; Governato et~al. 1998; de Lucia et~al. 2006; Benson 2010;
Alexander \& Hickox 2012). These distant high-density regions can be
identified as protoclusters (e.g.,\ Governato et~al. 1998; Chiang
et~al. 2013), structures where gravitational collapse and coalescence
has not yet been sufficient to produce a virialised galaxy
cluster. Direct comparisons between the observed growth rates of
galaxies and BHs in protoclusters to systems in the field can reveal
whether galaxy and BH growth is accelerated in distant high-density
regions of the Universe.

One of the best studied high-density regions in the distant Universe
is the $z\approx$~3.09 protocluster in the SSA~22 field. The SSA~22
protocluster was originally identified as a significant overdensity
(factor $\approx$~4--6 when compared to the field) of Lyman-Break
Galaxies (LBGs) and is predicted to evolve into a
$\approx10^{15}$~$M_{\odot}$ galaxy cluster (i.e.,\ similar to the
Coma cluster) by the present day (e.g.,\ Steidel et~al. 1998, 2000;
Kubo et~al. 2015). The protocluster has been traced over a
60~$\times$~10~Mpc$^2$ (co-moving) region using narrow-band imaging at
rest-frame Ly-$\alpha$ wavelengths, which also reveals a significant
overdensity of Ly-$\alpha$ Emitters (LAEs; e.g.,\ Hayashino
et~al. 2004; Matsuda et~al. 2005; Yamada et~al. 2012b) and regions of
extended ($\simgt$~30~kpc) Ly-$\alpha$ emission (often termed
Lyman-Alpha Blobs, LABs; e.g.,\ Steidel et~al. 2000; Matsuda
et~al. 2004; Yamada et~al. 2012b) when compared to the field. The
SSA~22 protocluster therefore provides an ideal environment to test
whether galaxy and BH growth is accelerated in a distant high-density
environment.

Deep {\it Chandra} observations of the SSA~22 protocluster have
revealed a significant enhancement (factor $6.1^{+10.3}_{-3.6}$) in
the fraction of galaxies that host AGN activity above a given X-ray
luminosity threshold when compared to the field at $z\approx$~3
(Lehmer et~al. 2009a). The increase in the AGN fraction indicates an
increase in the duty cycle of AGN activity (i.e.,\ the duration of
significant BH growth rates) over that found in the field. Given the
broad connection between AGN activity and star formation and the
various tight relationships between BH mass and the properties of
nearby galaxies (e.g.,\ Alexander \& Hickox 2012; Graham \& Scott
2013; Kormendy \& Ho 2013), we may also expect an enhancement in the
amount of star formation per galaxy in the SSA~22 protocluster. This
hypothesis can be tested by measuring the SFRs of the AGNs and
galaxies in the protocluster. However, with the exception of the deep
ALMA observations of the protocluster core by Umehata et~al. (2015),
the existing SFR measurements for the SSA~22 protocluster have been
taken at wavelengths where the emission is either easily obscured by
dust (e.g.,\ the ultra-violet continuum or Ly~$\alpha$ emission;
Matsuda et~al. 2005) or the data is too shallow to provide sensitive
individual SFR constraints on all but the brightest sources
(single-dish far-infrared--millimetre observations; e.g.,\ Geach
et~al. 2005, 2014; Scott et~al. 2006; Tamura et~al. 2009, 2010, 2013;
Umehata et~al. 2014).

In this paper we present ALMA 870~$\mu$m observations of eight X-ray
detected AGNs in the SSA~22 protocluster to provide sensitive SFR
measurements from even the most heavily obscured star-forming regions
in these sources. The main objective of this paper is to provide
sensitive constraints on the SFRs of AGNs in a distant protocluster
environment and to assess whether the host galaxies are growing more
rapidly than co-eval AGNs in the field. Four of the eight X-ray AGNs
are also coincident with LABs and we use our data to provide insight
on the star-formation properties of LABs. We have adopted
$H_{0}=71\kms$, $\Omega_{M}=0.27$ and $\Omega_{\Lambda}=0.73$; in this
cosmology 1$''$ corresponds to 7.8\,kpc at $z=3.09$. In our SFR
calculations we have assumed the Chabrier (2003) initial mass
function.


%
\section{Data}
\label{sec:obs_red}

\subsection{Sample Selection}

We selected our eight ALMA targets from the {\it Chandra} catalogues
of Lehmer et~al. (2009a,b) in the SSA~22 field. All of the targets
have spectroscopic redshifts of $z=$~3.08--3.11, placing them well
within the redshift range of the SSA~22 protocluster (Matsuda
et~al. 2005). All of the targets are also luminous at X-ray energies,
with X-ray luminosities of $L_{\rm 2-32
  keV}$~$=$~(0.9--4.2)~$\times10^{44}$~erg~s$^{-1}$, indicating that
they are X-ray AGNs.\footnote{We note that the rest-frame 2--32~keV
  luminosity is a factor of 2.2 larger than the more commonly used
  rest-frame 2--8~keV luminosity, assuming a typical X-ray spectral
  slope of $\Gamma=1.9$.} The high rest-frame energies probed by the
{\it Chandra} data at $z\approx$~3.09 (2--32~keV) also mean that all
but the most heavily obscured luminous AGNs (i.e.,\ $N_{\rm
  H}\simlt10^{24}$~cm$^{-2}$) will be detected and identified in the
X-ray band (e.g.,\ Alexander et~al. 2008; Brandt \& Alexander 2015;
Del Moro et~al. 2016). We have not attempted to correct the X-ray
luminosities for obscuration since the corrections will only be
significant for the most heavily obscured systems and would require
detailed X-ray spectral analyses and higher-quality X-ray data than
currently available to measure accurate column densities.\footnote{For
  example, corrections to the rest-frame 2--32~keV luminosity due to
  obscuration are only a factor $\simgt$~2 when $N_{\rm
    H}\simgt8\times10^{23}$~cm$^{-2}$ for a typical X-ray spectral
  slope of $\Gamma=1.9$.}

\begin{figure}
\includegraphics[width=8.0cm,angle=0]{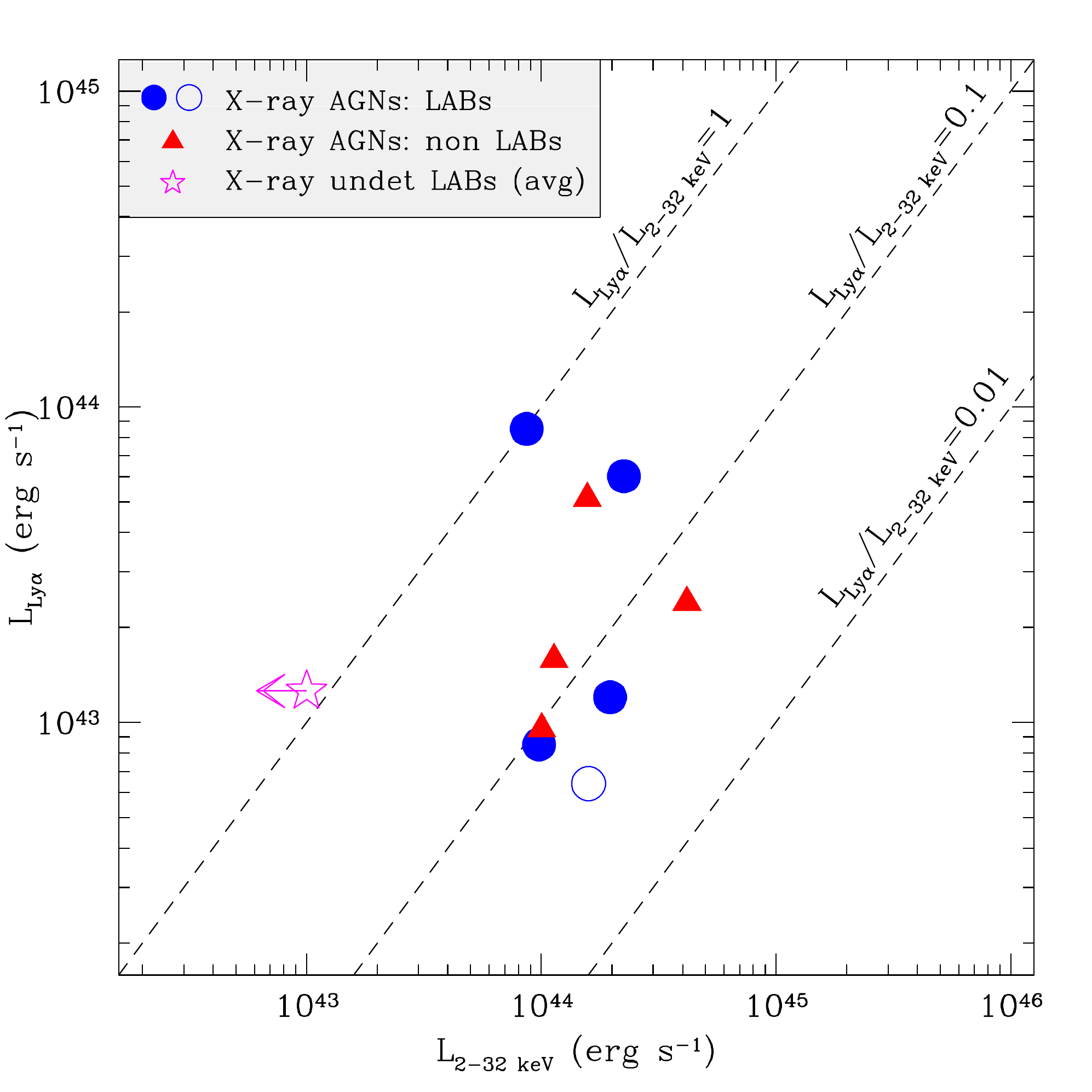}
\caption{Ly~$\alpha$ luminosity versus rest-frame 2--32~keV luminosity
  of the eight ALMA targets (filled circle: LABs; filled triangle: non
  LABs), and LAB~18 (open circle). As a comparison the average
  properties of the X-ray undetected LABs in the protocluster are also
  shown (open star). The dashed lines indicate constant ratios of
  Ly~$\alpha$ and X-ray luminosity. All of the X-ray AGNs are luminous
  ($L_{\rm X}\simgt10^{44}$~erg~s$^{-1}$) and, overall, they cover a
  broad range in Ly~$\alpha$/X-ray luminosity ratio.}
\label{fig:N24_onedspec}
\end{figure}

\begin{figure*}
\includegraphics[width=18.0cm,angle=0]{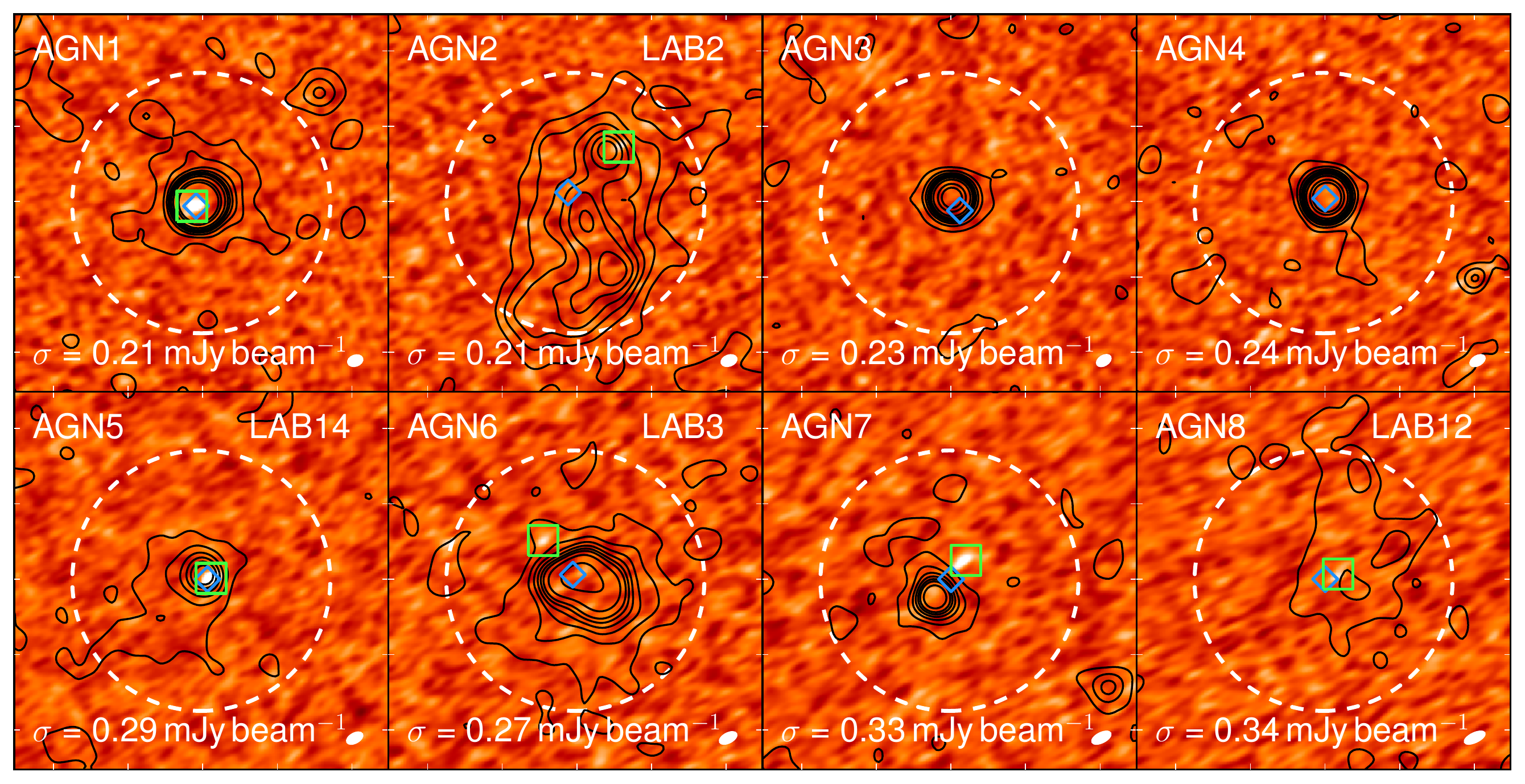}
\caption{ALMA 870~$\mu$m maps of the eight X-ray detected AGNs. The
  black contours trace the surface brightness of the Ly~$\alpha$
  emission, the blue diamond indicates the position of the X-ray AGN,
  the green square indicates the position of the ALMA-detected source,
  and the dashed circle indicates the size of the primary beam
  ($r=8.7$$''$), which corresponds to a projected radius of 68~kpc at
  $z=$~3.09. The size and shape of the synthesized beam is shown in
  the bottom right-hand corner of each map. We find an ALMA
  counterpart directly matched to an X-ray AGN in three sources
  (AGN~1; AGN~5; AGN~8) and an ALMA counterpart offset by 1.6--4.5$''$
  from an X-ray AGN in a further three sources (AGN~2; AGN~6;
  AGN~7). An ALMA counterpart lies within the extent of the
  Ly~$\alpha$ emission for all of the four LABs. See Tables 1 \& 2 for
  the source properties.}
\label{fig:N24_onedspec}
\end{figure*}

Our sample includes all six of the X-ray AGNs identified at
$z\approx$~3.09 in Lehmer et~al. (2009a) plus two additional X-ray
AGNs from Lehmer et~al. (2009b) that have been spectroscopically
identified as lying at $z\approx$~3.09 (AGN~7: $z=3.098$; Saez
et~al. 2015; AGN~8: $z=3.091$; Kubo et~al. 2015). This sample also
contains four of the five X-ray detected LABs in Geach
et~al. (2009). However, there are several X-ray AGNs in the
protocluster that we did not observe with ALMA (Tamura et~al. 2010;
Kubo et~al. 2013; Umehata et~al. 2015), including LAB~18 from Geach
et~al. (2009). On the basis of optical spectral analyses (Steidel
et~al. 2003; Yamada et~al. 2012a; Kubo et~al. 2015), five of the X-ray
AGNs in our sample have narrow emission lines and therefore appear to
be optically obscured AGNs while three have broad emission lines and
are therefore optically unobscured AGNs. Four of the X-ray AGNs are
identified as LABs (LAB~2; LAB~3; LAB~12; LAB~14), two of which have
physical extents of $\simgt$~100~kpc and are defined as giant LABs
(LAB~2; LAB~3; Matsuda et~al. 2011).

\begin{table*}
{\scriptsize
\begin{center}
{\centerline{\sc ALMA Properties of the X-ray detected AGNs}}
\smallskip
\begin{tabular}{lcccccccccc}
\hline
\noalign{\smallskip}
Name  & $\alpha_{\rm Chandra}$ & $\delta_{\rm Chandra}$ & $z$ & $L_{\rm 2-32 keV}$  & $L_{\rm Ly~\alpha}$ & $\alpha_{\rm ALMA}$ & $\delta_{\rm ALMA}$ & Offset   & $S_{\rm 870{\mu m}}$ & $L_{\rm IR}$ \\
       & (J2000)  & (J2000) &   & (log(erg~s$^{-1}$)) & (log(erg~s$^{-1}$)) & (J2000)  & (J2000)  & (arcsec) & (mJy) & (log(erg~s$^{-1}$)) \\
\hline
AGN~1           & 22 17 36.54 & +00 16 22.6 & 3.084 & 44.20 & 43.71 & 22 17 36.56 & +00 16 22.6 & 0.27  & $1.84\pm0.21$ & $45.98$ \\
AGN~2 (LAB~2)   & 22 17 39.08 & +00 13 30.7 & 3.091 & 43.94 & 43.93 & \dots       & \dots       & \dots & $<0.92$       & $<45.68$ \\
AGN~3           & 22 17 09.60 & +00 18 00.1 & 3.106 & 44.05 & 43.20 & \dots       & \dots       & \dots & $<1.04$       & $<45.73$ \\
AGN~4           & 22 17 20.24 & +00 20 19.3 & 3.105 & 44.62 & 43.38 & \dots       & \dots       & \dots & $<1.10$       & $<45.76$ \\
AGN~5 (LAB~14)  & 22 17 35.84 & +00 15 59.1 & 3.094 & 44.29 & 43.08 & 22 17 35.82 & +00 15 59.2 & 0.25  & $2.96\pm0.29$ & $46.19$ \\
AGN~6 (LAB~3)   & 22 17 59.23 & +00 15 29.7 & 3.096 & 44.35 & 43.78 & \dots       & \dots       & \dots & $<1.21$       & $<45.80$ \\
AGN~7           & 22 17 16.16 & +00 17 45.8 & 3.098 & 44.00 & 42.98 & \dots       & \dots       & \dots & $<1.50$       & $<45.89$ \\
AGN~8 (LAB~12)  & 22 17 32.00 & +00 16 55.6 & 3.091 & 43.99 & 42.93 & 22 17 31.94 & +00 16 55.9 & 0.91  & $1.58\pm0.35$ & $45.92$ \\
\hline\hline
\label{table:SMMJ1237_multispec}
\end{tabular}
\caption{The coordinates correspond to the {\it Chandra} source
  position and the ALMA source position and the offset refers to the
  angular separation between the {\it Chandra} and ALMA source. The
  redshifts are all spectroscopic and are taken from Steidel
  et~al. (2003), Matsuda et~al. (2005), Kubo et~al. (2015), and Saez
  et~al. (2015). The X-ray luminosity is calculated at rest-frame
  2--32~keV using the 0.5--8~keV flux from Lehmer et~al. (2009b) and
  the Ly~$\alpha$ luminosity is calculated using the Ly~$\alpha$ flux
  from Matsuda et~al. (2004); see footnote~1 for the X-ray luminosity
  convection to rest-frame 2--8~keV. The ALMA 870~$\mu$m measurements
  (source flux and uncertainty) are primary-beam corrected values and
  the upper limits refer to 4.5 times the primary-bream corrected
  rms. The infrared luminosity ($L_{\rm IR}$) refers to the
  star-formation emission at rest-frame 8--1000~$\mu$m and has an
  uncertainty of $\approx$~0.3~dex; see \S2.4 for the calculation of
  $L_{\rm IR}$.}
\end{center}
}
\end{table*}

All of our targets are detected in the narrow-band rest-frame
Ly~$\alpha$ imaging of Matsuda et~al. (2004). In Fig.~1 we plot the
Ly~$\alpha$ luminosity versus the X-ray luminosity of the ALMA
targets. The mean X-ray and Ly~$\alpha$ luminosities of the LABs and
non LABs are comparable: log($L_{\rm 2-32
  keV}$/erg~s$^{-1}$)~=~$44.1\pm0.2$ and log($L_{\rm
  Ly\alpha}$/erg~s$^{-1}$)~=~$43.4\pm0.5$ for the LABs and log($L_{\rm
  2-32 keV}$/erg~s$^{-1}$)~=~$44.1\pm0.2$ and log($L_{\rm
  Ly\alpha}$/erg~s$^{-1}$)~=~$43.3\pm0.3$ for the non LABs. However,
overall the ALMA targets cover a broad range in Ly~$\alpha$/X-ray
luminosity ratio. The Ly~$\alpha$/X-ray luminosity ratio provides
insight on the AGN contribution to the Ly~$\alpha$ emission, which is
discussed in \S3.2. We note that the distinction between LABs and non
LABs is not based on Ly~$\alpha$ luminosity but is a function of both
the extent and surface brightness of the Ly~$\alpha$ emission; see \S3
of Matsuda et~al. (2004). The mean Ly~$\alpha$ luminosity of the X-ray
undetected LABs in the SSA~22 protocluster (log($L_{\rm
  Ly\alpha}$/erg~s$^{-1}$)~=~$43.1\pm0.3$) is also similar to the ALMA
targets despite being at least an order of magnitude fainter at X-ray
energies ($L_{\rm 2-32 keV}\simlt10^{43}$~erg~s$^{-1}$; Geach
et~al. 2009).

\subsection{ALMA Observations and Data Reduction}

The eight targets were observed with ALMA on 20th November 2012, as
part of the Cycle~0 project 2011.0.00725. Each target was observed
using a 7.5~GHz bandwidth, centred on 344~GHz (i.e.,\ 870~$\mu$m:
band~7). A single--continuum correlator set-up was used, with four
basebands of 128 dual-polarization channels each. The array
configuration was such that a total of 25 antennae were used, with a
maximum baseline of 375\,m and a median baseline of 145\,m. The
minimum baseline of the array is 15\,m, which translates to a maximum
recoverable size of $\approx$~7$''$.

Each target was observed for a total of 310\,s. Neptune was used as
the primary flux calibrator, with J\,2225$-$049 used for band pass and
phase calibration. Neptune is known to have a CO absorption feature at
345~GHz. To minimise the impact of this feature on our flux
measurements we have modelled the data using the
2012-Butler-JPL-Horizons model for Neptune, which includes CO
absorption lines. Furthermore, our ALMA observations only have
coverage at 337--340~GHz and 350--353~GHz and, therefore, the CO
absorption feature will have a negligible impact on our measurements.

The data were processed with the Common Astronomy Software Application
({\sc casa} v4.4.0; McMullin et~al. 2007), and maps were produced
using the {\sc clean} routine within {\sc casa}. As is the standard
approach for interferometry data, we ``cleaned'' each target map to
reduce the strength of the side lobes from detected sources. This is
required to accurately measure the properties of all detected sources
and to search for faint sources that lie close to the side lobes of a
bright source. For each target we adopt an iterative approach to the
clean procedure, following the method outlined in Hodge et~al. (2013)
and Simpson et~al. (2015). We first create a ``dirty'' map of each
target, using natural weighting, and measure the root mean square
(rms; $\sigma)$ noise in the resulting map. Tight clean boxes were
then placed around all sources detected at $>$\,5\,$\sigma$ and the
dirty map is cleaned to a depth of 1.5\,$\sigma$ within these clean
boxes. We then measure the rms in this initial cleaned map, or use the
rms in the dirty map if no sources are detected at $>$\,5\,$\sigma$,
and repeat the clean procedure on all sources detected at
$>$\,3.5\,$\sigma$ in these maps to produce a final map that is
suitable for the detection of even faint sources. If no sources are
detected at $>$\,3.5\,$\sigma$ then the dirty map is considered the
final map.

The final ALMA maps for our eight targets have a range of rms values
($\sigma=$~0.21--0.34\,mJy~beam$^{-1}$) and a median synthesized beam
of 1.10$''$\,$\times$\,0.61$''$. We note that the synthesized beam
becomes increasingly elongated at low target elevations, and the final
maps have a range of beam major and minor axes of 0.99--1.45$''$ and
0.68--0.59$''$, respectively; see Fig.~2 for the size and shape of the
synthesized beam for each map. Each map was created with a total size
of 25.6$''$\,$\times$\,25.6$''$ and a pixel scale of 0.1$''$.

\begin{table}
{\scriptsize
\begin{center}
{\centerline{\sc Additional ALMA-detected Sources in the Protocluster}}
\smallskip
\begin{tabular}{lccccc}
\hline
\noalign{\smallskip}
Name  & $\alpha_{\rm ALMA}$ & $\delta_{\rm ALMA}$ & Offset & $S_{\rm 870{\mu m}}$ & $L_{\rm IR}$ \\
       & (J2000)  & (J2000) &  (arcsec) & (mJy) & (log(erg~s$^{-1}$)) \\
\hline
AGN~2 (LAB~2) & 22 17 38.85 & +00 13 33.7 & 4.54 & $1.11\pm0.25$ & 45.76 \\
AGN~6 (LAB~3) & 22 17 59.34 & +00 15 32.0 & 2.81 & $1.38\pm0.29$ & 45.86 \\
AGN~7 & 22 17 16.09 & +00 17 47.0 & 1.57 & $2.25\pm0.34$ & 46.07 \\
\hline\hline
\label{table:SMMJ1237_multispec}
\end{tabular}
\caption{The coordinates correspond to the ALMA source position and
  the offset refers to the angular separation between the ALMA and
  {\it Chandra} source; see Table~1 for the {\it Chandra} source
  positions. The ALMA 870~$\mu$m measurements (source flux and
  uncertainty) are primary-beam corrected values. The infrared
  luminosity ($L_{\rm IR}$) refers to the star-formation emission over
  rest-frame 8--1000~$\mu$m, calculated assuming $z=3.09$, and has an
  uncertainty of $\approx$~0.3~dex; see \S2.4 for the calculation of
  $L_{\rm IR}$.}
\end{center}
}
\end{table}

\subsection{ALMA Source Detection, Matching, and Properties}

The ALMA maps for the eight targets are shown in Fig.~2, with contours
of the Ly~$\alpha$ emission overlaid. Several apparently significant
peaks at 870~$\mu$m are seen in the maps but we need to set a
detection threshold to reliably distinguish between real and spurious
sources. To achieve this we initially identified all $>$3.5\,$\sigma$
peaks in the non-primary-beam corrected ALMA maps as potential sources
and then inverted the maps and repeated this detection procedure. To
measure the spurious detection rate for a given significance threshold
we then simply compare the number of detected sources between the
original and inverted maps as a function of the detection
threshold. We find that the number of ``negative'' sources falls to
zero at $>$\,4.5\,$\sigma$, and hence to ensure that we only include
robust ALMA detections we only consider peaks in the ALMA maps at
$>$\,4.5\,$\sigma$. Overall we detect six ALMA sources at
$>$\,4.5\,$\sigma$ within the primary beam of the ALMA maps; see
Fig.~2. We measured the peak flux density and fitted a point source
model at the position of each source to search for evidence of
extended 870~$\mu$m emission. We do not see significant residuals
after subtracting the best-fit point source model, indicating that the
sources are not resolved in our maps and corresponding to physical
scales for the star-formation emission of $\simlt$~11~kpc. The peak
flux density in the map is therefore taken to be the flux density of
each source, which we correct for the primary beam attenuation.

We searched for matches between the X-ray AGNs and the ALMA sources
using the {\it Chandra} and ALMA positional uncertainties. For the
uncertainty in the X-ray source position we used 1.5 times the 80\%
confidence level of each X-ray source from Lehmer et~al. (2009b) and
for the uncertainty in the ALMA source position we conservatively
assumed 0.4$''$, the average ALMA positional uncertainty measured by
Hodge et~al. (2013). Combining these two positional uncertainties we
find that three of the six ALMA sources are directly matched with an
X-ray AGN (AGN~1; AGN~5; AGN~8; see Table~1), with {\it Chandra}--ALMA
positional offsets of 0.3--0.9$''$. The other three ALMA sources have
1.6--4.5$''$ offsets from {\it Chandra} sources and are not directly
matched to an X-ray AGN (see Table~2). The 1.6$''$ offset
($\approx$~12~kpc in projection) between the X-ray AGN and ALMA source
in AGN~7 is larger than our search radius but still close enough that
the two sources may be physically associated. Indeed, investigation of
publicly available optical--near-IR {\it Hubble Space Telescope}
imaging shows that the X-ray AGN is matched to a point source while
the ALMA source is matched to a galaxy with a disturbed morphology,
suggesting that these are two distinct sources (an X-ray AGN and a
galaxy) in a merger with a projected separation of
$\approx$~12~kpc. We note that a larger fraction of galaxies appear to
reside in mergers in the SSA~22 protocluster when compared to coeval
galaxies in the field (Hine et~al. 2016a).

The two ALMA sources in the fields of AGN~2 and AGN~6 have large {\it
  Chandra}--ALMA offsets of 2.8-4.5$''$ ($\approx$~22--35~kpc in
projection) and are not directly matched to an X-ray AGN. However,
each ALMA source resides within the extended Ly~$\alpha$ emission of a
LAB (LAB~2; LAB~3). This suggests that the ALMA sources may be
physically associated with the LAB; see Fig.~2. We can provide a basic
test of this scenario by assessing whether we would expect to detect
any sources by chance in our ALMA maps. To determine the number of
sources expected in our maps by chance we took the best-fitting model
of the 870~$\mu$m number counts from Simpson et~al. (2015) and
calculated the number of ALMA sources that we would expect in each
map, taking into account both the sensitivity of each map and the
decrease in sensitivity from the phase centre. On the basis of this
simple test, overall, we predict 0.17 ALMA sources by chance within
the primary beam across all of our eight ALMA maps and 0.04 ALMA
sources within the primary beam across the two ALMA maps of LAB~2 and
LAB~3, suggesting that the offset ALMA sources are likely to be
physically associated with the LAB.

The 870~$\mu$m flux densities of the six ALMA-detected sources are
1.10--2.96~mJy. We calculated ALMA upper limits for the five X-ray
AGNs without an ALMA counterpart by taking 4.5 times the rms,
adjusting for any small primary beam corrections when the X-ray source
does not lie at the phase centre. The ALMA properties of the eight
X-ray AGNs are given in Table~1 and the ALMA properties for the three
additional ALMA sources not directly matched to an X-ray AGN are given
in Table~2. The only source in our sample reliably detected at
submillimeter wavelengths in previous studies is the brightest ALMA
source (AGN~5), which has a flux density from SCUBA observations
consistent within $\approx$~1.5~$\sigma$ of the ALMA flux density
($S_{\rm 850~{\mu}m}=4.9\pm1.3$~mJy; Geach et~al. 2005). Two of the
other X-ray AGNs have been previously detected by ALMA at 1.1~mm
wavelengths (AGN~1; AGN~8; Umehata et~al. 2015) and our 870~$\mu$m
flux densities for these two sources are $\approx$~1.8--2.3 times
higher than the 1.1~mm flux densities measured by Umehata
et~al. (2015), which is within the range expected for dust emission
from a typical star-forming galaxy at this redshift. The flux
densities of the other three ALMA-detected sources are below the
sensitivity limits of previous-generation submillimetre and millimetre
observatories.

\subsection{Measurement of star-formation rates}

The rest-frame wavelengths of the ALMA data correspond to
$\approx$~210~$\mu$m at $z=3.09$. Such long-wavelength far-infrared
emission is likely to be dominated by star-formation activity, which
typically peaks at $\approx$~100~$\mu$m (e.g.,\ Brandl et~al. 2006;
Mullaney et~al. 2011; Bethermin et~al. 2015). By comparison, dust
emission from AGN activity peaks at $<40$~$\mu$m and drops off sharply
at longer wavelengths (e.g.,\ Richards et~al. 2006; Netzer
et~al. 2007; Mullaney et~al. 2011).

We calculated the infrared luminosities ($L_{\rm IR}$ over rest-frame
8--1000~$\mu$m) from star formation using the ALMA 870~$\mu$m flux
densities following \S3 of Mullaney et~al. (2015). We adopted this
specific approach since the Mullaney et~al. (2015) study provided SFR
measurements of X-ray AGNs in the field on the basis of ALMA
870~$\mu$m data, which are used as our field AGN comparison sample in
\S3. Briefly, the infrared luminosities are calculated from the
870~$\mu$m flux densities over rest-frame 8--1000~$\mu$m using the
source redshifts and the average SEDs of the star-forming galaxies in
Bethermin et~al. (2015). The uncertainty on $L_{\rm IR}$ is relatively
modest because the ALMA data probes close to the peak of the SED for
star-forming galaxies at $z\approx$~3 and is estimated by Mullaney
et~al. (2015) to be $\approx$~0.3~dex on the basis of the range of SED
templates of Draine \& Li (2007); see \S3 of Mullaney
et~al. (2015). See Tables 1 \& 2 for the calculated $L_{\rm IR}$
values. SFRs are estimated from $L_{\rm IR}$ following Kennicutt
(1998) for the Chabrier (2003) initial mass function; the conversion
from $L_{\rm IR}$ to SFR adopted in our study is

\begin{equation}
{\rm SFR} = {{L_{\rm IR}\over{3.778\times10^{43}~erg~s^{-1}}}}
M_{\odot}~yr{^{-1}}.
\end{equation}

The rest-frame 210~$\mu$m emission is likely to be dominated by star
formation for all but the most luminous AGNs. We can verify this and
quantify the potential contribution to the ALMA flux densities from
AGN activity by taking the {\it Spitzer} 24~$\mu$m constraints of our
sources and predicting the 870~$\mu$m flux density. All of our X-ray
AGNs and ALMA sources have a 24~$\mu$m flux density of $S_{\rm
  24~{\mu}m}<100$~$\mu$Jy, with the exception of AGN~1 which has
$S_{\rm 24~{\mu}m}=450\pm10$~$\mu$Jy (e.g.,\ Webb et~al. 2009; Colbert
et~al. 2011; Capak et~al. 2013). On the basis of the mean empirical
AGN spectral energy distribution (SED) template of Mullaney
et~al. (2011), a 24~$\mu$m flux density of 100~$\mu$Jy would
correspond to a 870~$\mu$m flux density of only 35~$\mu$Jy at
$z=3.09$. Conservatively assuming that the 24~$\mu$m emission is
dominated by AGN activity rather than star formation, we therefore
predict that the AGN contributes to $\approx$~9\% of the 870~$\mu$m
flux density for AGN~1 and contributes to $\simlt$~4\% for all of the
other sources. We therefore expect our 870~$\mu$m flux densities to
provide a reliable measurement of the star-formation luminosities of
our sources.

\section{Results}
\label{sec:SMMJ1237}

Overall we detected six sources at 870~$\mu$m within the primary beam
of the eight ALMA maps: three of the eight X-ray AGNs have ALMA
counterparts, all of the four LABs have an ALMA counterpart within the
extent of the Ly~$\alpha$ emission (two of which are directly matched
with an X-ray AGN and two of which are offset from the X-ray AGN but
are likely to be physically associated with the LAB), and one ALMA
source appears to be a galaxy in a merger with an X-ray AGN. In the
following sub sections we compare the SFRs of the X-ray AGNs in the
protocluster to the SFRs of distant X-ray AGNs in the field (see
\S3.1) and investigate the SFRs of the protocluster LABs (see \S3.2).

\subsection{The star-formation properties of X-ray detected AGNs in a distant protocluster}

\begin{figure*}
\includegraphics[width=10.5cm,angle=0]{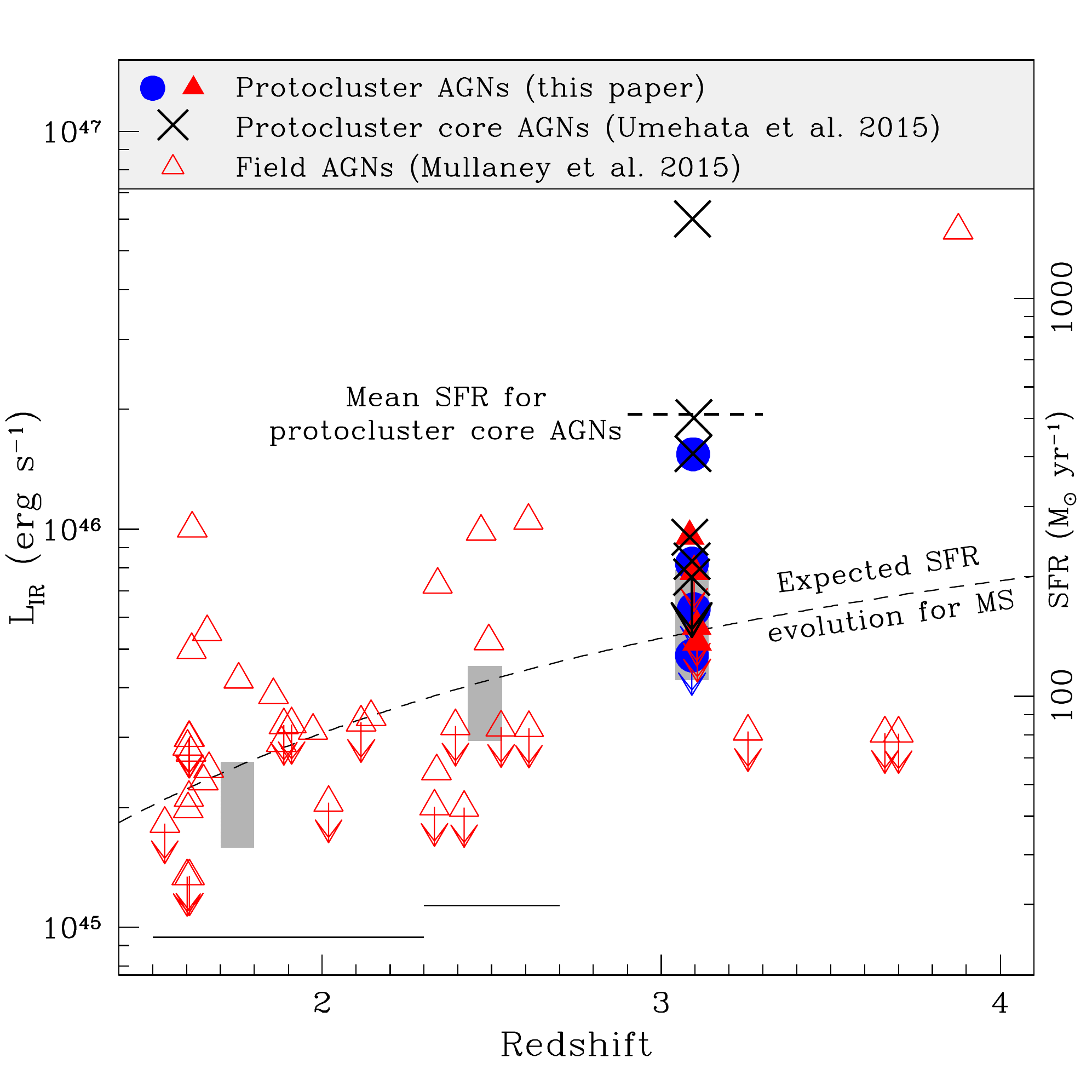}
\caption{Infrared luminosity from star formation versus redshift for
  the X-ray detected AGNs in the SSA~22 protocluster from this paper
  compared to X-ray detected AGNs in the protocluster core (crosses;
  taken from Umehata et~al. 2015) and X-ray detected AGNs in the field
  (open triangles; taken from Mullaney et~al. 2015); see Fig.~1 for
  the other symbol descriptions. The conversion from $L_{\rm IR}$ to
  SFR is calculated using Eqn.~1. The shaded regions indicate the
  range in mean infrared luminosities for the protocluster AGNs at
  $z\approx$~3.09 (excluding the Umehata et~al. 2015 data) and for the
  AGNs in the field at two mean redshifts; the redshift ranges are
  indicated by the black horizontal lines. The widths of the shaded
  regions are arbitrary. The dashed line indicates the mean infrared
  luminosity for the protocluster AGNs at the core (using the Umehata
  et~al. 2015 data) and the dashed curve indicates the measured
  evolution in infrared luminosity for an $M_{*}$ star-forming galaxy
  over $z=$~1.4--4.1 from Schreiber et~al. (2015), using the
  definition of $M_{*}$ from Ilbert et~al. (2013). There is
  significant scatter in the infrared luminosities for individual
  X-ray AGNs; however, the mean infrared luminosities are consistent
  with $M_{*}$ star-forming galaxies at the average redshifts for both
  the protocluster AGN from this paper and the field AGN
  samples. Including the Umehata et~al. (2015) data shows that the
  infrared luminosities are significantly enhanced for the AGNs at the
  protocluster core.}
\label{fig:N24_onedspec}
\end{figure*}

Many previous studies have explored the SFRs of distant X-ray AGNs in
the field (e.g.,\ Shao et~al. 2010; Harrison et~al. 2012; Mullaney
et~al. 2012; Rosario et~al. 2012, 2013; Santini et~al. 2012; Stanley
et~al. 2015). However, little is known about the SFRs of X-ray AGNs in
a distant protocluster environment. The previous studies of field AGNs
have found that both the mean SFR and the evolution in SFR with
redshift are consistent with those of co-eval massive
($\approx$~$M_{*}$) star-forming galaxies (e.g.,\ Mullaney
et~al. 2012; Santini et~al. 2012; Hickox et~al. 2014; Stanley
et~al. 2015); for example, the mean SFRs of X-ray AGNs at
$z\approx$~2.0--2.5 are $\approx$~10 times higher than those of X-ray
AGNs at $z\approx$~0.5--1.0. On average, distant X-ray AGNs therefore
appear to reside in typical star-forming galaxies, although we caution
that there can be a broad spread of individual SFR values
(e.g.,\ Mullaney et~al. 2015).

In our analyses we compare the SFRs of X-ray AGNs in the protocluster
and field, utilising the study of Mullaney et~al. (2015), which
primarily used ALMA 870~$\mu$m data to measure SFRs for $z>1.5$ X-ray
AGNs in the field. The field AGNs extend to lower X-ray luminosities
than the protocluster AGNs, as expected due to the deeper {\it
  Chandra} data from Xue et~al. (2011) utilised in Mullaney
et~al. (2015). The rest-frame 2--8~keV luminosities of the field and
protocluster AGNs are $L_{\rm 2-8
  keV}\approx10^{42}$--$2\times10^{44}$~erg~s$^{-1}$ and $L_{\rm 2-8
  keV}\approx4\times10^{43}$--$2\times10^{44}$~erg~s$^{-1}$,
respectively; we have converted the X-ray luminosities of the field
and protocluster AGNs to a common rest-frame 2--8~keV luminosity
assuming a typical X-ray spectral slope of $\Gamma=1.9$ (see
footnote~1). However, the lack of lower-luminosity AGNs in the
protocluster sample should not affect our comparison since the average
SFRs of AGNs are not a function of X-ray luminosity over the redshift
and luminosity ranges of our sources, at least for field AGNs
(e.g.,\ Mullaney et~al. 2012; Rosario et~al. 2012; Stanley
et~al. 2015). Since the SFR is broadly proportional to the galaxy mass
in star-forming galaxies (i.e.,\ what is often referred to as the
``main sequence''; e.g.,\ Elbaz et~al. 2011; Speagle et~al. 2014;
Schreiber et~al. 2015), it is also useful to consider the stellar
masses of the AGN host galaxies. The stellar masses of the field AGNs
in Mullaney et~al. (2015) have been calculated by Santini
et~al. (2012) from fitting AGN and host-galaxy templates to the
rest-frame optical--near-IR data. The stellar masses cover
(0.2--3)~$\times$~$10^{11}$~$M_{\odot}$ and the mean stellar mass is
$\approx5\times10^{10}$~$M_{\odot}$, which corresponds to $M_{*}$ over
the redshift range of the field AGNs (e.g.,\ Ilbert et~al. 2013). The
stellar masses of the protocluster AGNs have been calculated by Kubo
et~al. (2015) from fitting host-galaxy templates to the rest-frame
optical--near-IR data. Excluding the three protocluster AGNs with
broad optical emission lines (AGN~1; AGN~4; AGN~7; Steidel
et~al. 2003; Yamada et~al. 2012a; Kubo et~al. 2015), which indicate
the presence of an unobscured AGN at optical wavelengths that will
contaminate the host-galaxy mass measurements, the range in stellar
masses is (0.3--2)~$\times$~$10^{11}$~$M_{\odot}$, with a mean of
$\approx6\times10^{10}$~$M_{\odot}$. On the basis of this stellar mass
comparison there are no significant differences between the field and
protocluster AGNs, although we note that the protocluster AGN sample
is small and that there are significant uncertainties on the stellar
masses of individual sources.

In Fig.~3 we plot the infrared luminosity from star formation versus
redshift for the X-ray AGNs in the protocluster and the field. Using
Eqn~1 to convert from $L_{\rm IR}$ to SFR, the SFRs of the three
ALMA-detected X-ray AGNs are $\approx$~220--410~$M_{\odot}$~yr$^{-1}$
and the SFR upper limits for the five ALMA-undetected X-ray AGNs are
$\simlt$~210~$M_{\odot}$~yr$^{-1}$ (the SFR upper limits range from
$<130$~$M_{\odot}$~yr$^{-1}$ to $<210$~$M_{\odot}$~yr$^{-1}$). These
SFRs are broadly similar to those calculated for the X-ray AGNs in the
field, although to make a more quantitative comparison we need to
calculate mean SFRs. Many of the X-ray AGNs in the protocluster and
field have SFR upper limits, precluding the calculation of a mean, and
we have therefore adopted a simple approach to calculate the range in
mean SFR that covers all possibilities. The lower limit on the mean
SFR is calculated assuming that the SFRs of the sources with SFR upper
limits have the extreme value of 0~$M_{\odot}$~yr$^{-1}$ while an
upper limit is calculated by assuming that the SFRs are set at the
upper limit values. This approach is conservative since the true mean
SFR for each of the samples must lie within the calculated
ranges. Using this approach we calculated the following ranges in mean
SFR: 110--210~$M_{\odot}$~yr$^{-1}$ for the protocluster AGNs,
80--120~$M_{\odot}$~yr$^{-1}$ for the $z=$~2.3--2.7 field AGNs (mean
$z=$~2.48), and 40--70~$M_{\odot}$~yr$^{-1}$ for the $z=$~1.5--2.3
field AGNs (mean $z=$~1.75). A simple comparison of these ranges in
mean SFR shows that the protocluster AGNs have elevated SFRs over AGNs
in the field. However, to provide a more accurate comparison we must
also take into account the expected evolution in SFR with redshift of
the field AGNs out to the higher redshift of the protocluster.

The SFRs of X-ray AGNs in the field are found to track the evolution
of massive star-forming galaxies (galaxies with masses around $M_{\rm
  *}$, the knee of the stellar-mass function; e.g.,\ Mullaney
et~al. 2012; Santini et~al. 2012; Hickox et~al. 2014; Stanley
et~al. 2015) and therefore when accounting for the expected evolution
in SFR with redshift of the field AGNs we have assumed the measured
SFR evolution for $M_{\rm *}$ star-forming galaxies from Schreiber
et~al. (2015); we recall that the mean stellar mass of the
protocluster AGNs is also consistent with $M_{\rm *}$. On the basis of
this approach we expect a factor $\approx$~2.2 enhancement in mean SFR
from $z=1.75$ to $z=3.09$ (giving a predicted range of
90--150~$M_{\odot}$~yr$^{-1}$ for the field AGNs at $z=3.09$) and a
factor $\approx$~1.3 enhancement in mean SFR from $z=2.48$ to $z=3.09$
(giving a predicted range of 100--160~$M_{\odot}$~yr$^{-1}$ for the
field AGNs at $z=3.09$). Taking account of this assumed global
evolution in SFR, the mean ranges in SFR between the X-ray AGNs in the
field and protocluster are now broadly consistent; see
Fig.~3. Assuming the lowest and highest values in the mean SFR ranges
for the field and protocluster AGNs, the mean SFRs of the protocluster
AGNs are enhanced by a factor of $\approx$~0.7--2.3 over the mean SFRs
of the field AGNs.

Overall our results indicate that the growth rates of individual AGN
host galaxies in a protocluster environment are not significantly
elevated over those of AGNs in the field. The masses and SFRs of both
the protocluster and field AGNs are also similar to those of typical
massive ($M_{\rm *}$) star-forming galaxies; see Fig.~3. However,
consideration of the spatial location of the ALMA-detected AGNs in the
protocluster indicates that there may be an environmental dependence
on the mean SFRs. The three ALMA-detected AGNs in our sample (AGN~1;
AGN~5; AGN~8) lie within the core of the protocluster, at the
intersection of three filamentary structures traced by LAEs (Matsuda
et~al. 2005; see Fig.~3 of Umehata et~al. 2015). Umehata et~al. (2015)
mapped this central region over 1.5$^{\prime}$~$\times$~3.0$^{\prime}$
($\approx$~0.7~$\times$~1.4~Mpc) with ALMA at 1.1~mm and measured a
$\approx$~2 orders of magnitude increase in the SFR density in this
region of the protocluster when compared to the field. It is therefore
potentially significant that the three ALMA-detected AGNs in our
sample lie within this high SFR region. These ALMA-detected AGNs were
also detected at 1.1~mm by Umehata et~al. (2015), in addition to
another two X-ray AGNs spectroscopically identified to lie in the
protocluster (a third X-ray AGN with a photometric redshift consistent
with the protocluster redshift was undetected at 1.1~mm). How much
does the mean SFR change if we include these three additional X-ray
AGNs that lie in the protocluster core? Allowing for the 0.1~dex
increase in $L_{\rm IR}$ between the star-forming galaxy templates
adopted in our study and those adopted in Umehata et~al. (2015), the
range in mean SFR for the 11 X-ray AGNs is
270--360~$M_{\odot}$~yr$^{-1}$, leading to a factor of
$\approx$~1.7--4.0 enhancement in mean SFR over that measured in the
field. However, we note that the majority of the increase in mean SFR
is driven by a single extreme object which has a SFR that is
$\simgt$~3 times larger than any of the other X-ray AGNs (ADF~22a in
Umehata et~al. 2015 with an implied SFR of 1600~$M_{\odot}$~yr$^{-1}$
for the templates adopted in our study). ADF~22a is the brightest
submillimetre galaxy in the SSA~22 protocluster (Umehata et~al. 2014)
and appears to lie at the bottom of the protocluster gravitational
potential and may be the progenitor of the brightest cluster galaxy
(e.g.,\ Tamura et~al. 2010; Umehata et~al. 2015).

These results indicate that, while there is a lack of significant
enhancement in SFR for the protocluster AGNs in our initial sample,
the star formation in the highest-density region within the core of
the protocluster is significantly enhanced. For example, the range in
mean SFR for just the six X-ray AGNs in the protocluster core mapped
by Umehata et~al. (2015) is 500--530~$M_{\odot}$~yr$^{-1}$, a factor
of $\approx$~3.1--5.9 enhancement in mean SFR over that measured in
the field. However, these measurements are significantly elevated by
ADF~22a; removing ADF~22a from the sample gives a range in mean SFR of
280--320~$M_{\odot}$~yr$^{-1}$ and a factor of $\approx$~1.8--3.6
enhancement in SFR over the field. We also note that Lehmer
et~al. (2009b) found evidence that the enhancement in the fraction of
galaxies hosting AGN activity found by Lehmer et~al. (2009a) is
highest in the densest regions of the protocluster, indicating an
increase in the duty cycle of AGN activity which may be connected to
the elevated star formation found here. To provide a more accurate
assessment of the star formation in the protocluster and its
connection to AGN activity would require sensitive SFR measurements
for all of the AGNs and the galaxies across the extent of the
protocluster, which would also allow for detailed BH--galaxy growth
comparisons as a function of local galaxy density (e.g.,\ Lehmer
et~al. 2009a, 2013; Umehata et~al. 2015).

\subsection{Star Formation associated with Lyman-Alpha Blobs}

Four of the X-ray AGNs studied here are hosted within LABs (LAB~2;
LAB~3; LAB~12; LAB~14), which provides the opportunity to investigate
the star-formation properties for a subset of the LAB population. The
origin of the extended Ly~$\alpha$ emission from LABs is a matter of
significant debate (e.g.,\ Fardal et~al. 2001; Geach et~al. 2005,
2009; Nilsson et~al. 2006; Dijkstra \& Loeb 2009; Faucher-Giguere
et~al. 2010; Cen \& Zheng 2013; Overzier et~al. 2013; Ao et~al. 2015;
Prescott et~al. 2015), with three leading possibilities: (1) the
cooling of pristine gas within the dark-matter halo, potentially fed
by so-called ``cold-gas streams'', (2) the heating of gas by AGN
activity (photoionisation or AGN-driven jets), and (3) the heating of
gas by star-formation processes (photoionisation by young stars or
supernovae-driven winds). Previous studies of the LABs in the SSA~22
protocluster have suggested that the extended Ly~$\alpha$ emission is
predominantly due to the heating of the gas rather than cooling
(e.g.,\ Bower et~al. 2004; Geach et~al. 2005, 2009, 2014; Wilman
et~al. 2005; Smith et~al. 2008; Webb et~al. 2009; Colbert et~al. 2011;
Hayes et~al. 2015). By selection, all of our ALMA-observed LABs host
X-ray AGNs which, as shown by Geach et~al. (2009), already provides a
potential source of heating for the Ly~$\alpha$ emission through
photoionisation. However, all of the four LABs also host an ALMA
source, either directly associated with the X-ray AGN or within the
extent of the Ly~$\alpha$ emission, which indicates the presence of
luminous star formation; see Table~1 and 2.

Using Eqn~1 to convert from $L_{\rm IR}$ to SFR, the range of SFRs for
the LABs is found to be $\approx$~150--410~$M_{\odot}$~yr$^{-1}$ and
the mean and standard deviation is
$240\pm110$~$M_{\odot}$~yr$^{-1}$. The range and mean of the SFRs are
consistent with those found for massive star-forming galaxies at
$z\approx$~3 ($\approx$~150~$M_{\odot}$~yr$^{-1}$ for $M_{\rm *}$;
e.g.,\ Schreiber et~al. 2015) and, therefore, the X-ray detected LABs
do not appear to have significantly elevated SFRs when compared to the
field; see Fig.~3. This mean SFR is also comparable to that implied
from the mean submillimetre flux found by stacking the low-resolution
single-dish submillimetre data of the most extended LABs in the SSA~22
protocluster (LABs~1--12; $\approx190\pm40$~$M_{\odot}$~yr$^{-1}$;
Hine et~al. 2016b), which are similar systems to the LABs explored
here. By comparison the mean SFR for all of the LABs in the SSA~22
protocluster is $\simgt$~2 times lower than that calculated here but
includes many more compact systems
($\approx80\pm30$~$M_{\odot}$~yr$^{-1}$; Hine
et~al. 2016b). Interestingly, the two giant LABs in our sample (LAB~2;
LAB~3), with physical extents of $\simgt$~100~kpc (Matsuda
et~al. 2011), do not host more luminous star formation than the two
smaller ALMA-detected LABs despite being an order of magnitude more
luminous in Ly~$\alpha$ emission. To first order this suggests that
star-formation activity has less of an affect on the production of the
extended Ly~$\alpha$ emission in the giant LABs than for the smaller
LABs.


We can quantify the potential contribution to the Ly~$\alpha$ emission
from photoionisation in LABs by taking the same approach as Geach
et~al. (2009) and calculate the ratio between the
200--912~\AA\ luminosity from star formation (i.e.,\ the ultra-violet
wavelengths where the photon energies are high enough to produce
Ly~$\alpha$ through photoionisation) and the Ly~$\alpha$
luminosity. In this calculation we take our infrared-derived
star-formation luminosities and convert them to
200--912~\AA\ luminosities assuming the star-forming galaxy template
adopted in Geach et~al. (2009), which provides a good characterisation
of the composite SED of the X-ray detected LABs in the SSA~22
protocluster. On the basis of this approach we find that the
luminosity at 200--912~\AA\ due to star formation is always at least
an order of magnitude higher than the Ly~$\alpha$ luminosity and can
be responsible for producing the Ly~$\alpha$ luminosity with the
following escape fractions of 200--912~\AA\ photons: $\approx$~5\%
(LAB~2), $\approx$~3\% (LAB~3), $\approx$~0.3\% (LAB~12), and
$\approx$~0.2\% (LAB~14). By comparison, the average escape fraction
for all of the LABs in the protocluster is $\approx$~2\%, based on the
mean SFR from Hine et~al. (2016b) and the mean Ly~$\alpha$ luminosity
for the LABs. The infrared luminosities of the LABs indicate that they
harbour dust-obscured star formation and we would therefore expect
only a small fraction of the 200--912~\AA\ photons to be able to
directly escape and photoionise the extended Ly~$\alpha$
emission. Geach et~al. (2009) estimated that the average escape
fraction for LABs is $\approx$~0.6\%; however, we note that, given the
factor $\approx$~5 decrease in the stacked submillimetre flux for the
LABs in the SSA~22 protocluster between the Geach et~al. (2009) study
and Hine et~al. (2016b), an average escape fraction of $\approx$~3\%
is more plausible. We also note that this is effectively a lower limit
on the average escape fraction since it assumes that the ultra-violet
emission is only due to star formation, when there is likely to also
be a contribution from AGN activity. Indeed, more accurate assessments
of the escape fraction based on high spatial resolution ALMA and {\it
  HST}-STIS observations indicate that the escape fraction varies
substantially across individual sources (likely due to the patchiness
of the obscured dust) but can reach values in excess of $\approx$~10\%
(J.~E.~Geach, in prep). Despite these caveats we note that our
calculated escape fractions are broadly consistent with those
estimated for star-forming galaxies at $z\approx$~3 (e.g.,\ Iwata
et~al. 2009; Siana et~al. 2015).

Therefore, on the basis of our results, star formation appears to be a
plausible mechanism to produce the Ly~$\alpha$ emission for all of the
LABs explored here, although the escape fraction of
200--912~\AA\ photons for LAB~2 and LAB~3 need to be comparatively
high. As calculated by Geach et~al. (2009), AGN activity also appears
to be sufficient to be able to produce the Ly~$\alpha$ luminosity for
all of the X-ray detected LABs in our sample (e.g.,\ on the basis of
the Ly~$\alpha$/X-ray luminosity ratio; see Fig.~4 in Geach
et~al. 2009). However, since the ALMA sources for LAB~2 and LAB~3 are
offset from the centre of the Ly~$\alpha$ emission and the X-ray AGN,
it is also possible that multiple systems photoionise the Ly~$\alpha$
emission in these giant LABs (as also potentially found for LAB~1;
Weijmans et~al 2010) and deeper ALMA observations may reveal fainter
star-forming galaxy components. We finally note that, since the ALMA
sources are unresolved in these four LABs (see \S2.3), this places
$\simlt$~11~kpc constraints on the physical scale of the
star-formation emission region.

\section{Conclusions}

We have presented ALMA 870~$\mu$m observations and calculated the SFRs
of eight X-ray detected AGNs, four of which reside within LABs, in the
$z\approx$~3.1 SSA~22 protocluster. With these data we have found the
following results:

\begin{itemize}

\item Three of the protocluster AGNs are detected by ALMA and have
  implied SFRs of $\approx$~220--410~$M_{\odot}$~yr$^{-1}$; the non
  detection of the other AGNs places SFR upper limits of
  $\simlt$~210~$M_{\odot}$~yr$^{-1}$. The mean SFR of the protocluster
  AGNs ($\approx$~110--210~$M_{\odot}$~yr$^{-1}$) is consistent
  (within a factor of $\approx$~0.7--2.3) with that found for co-eval
  AGNs in the field, implying that galaxy growth is not significantly
  accelerated across the protocluster environment. However, when also
  considering ALMA data from the literature, we find some evidence for
  significantly elevated mean SFRs (up-to a factor of $\approx$~5.9
  over the field) for the AGNs at the core of the protocluster,
  indicating that the mean growth of galaxies is accelerated in the
  central region. We note that the mean SFR at the protocluster core
  is significantly enhanced by a single extreme object, potentially
  the progenitor of the brightest cluster galaxy, with a SFR of
  $\approx$~1600~$M_{\odot}$~yr$^{-1}$.

\item All four of the protocluster LABs are associated with an ALMA
  source within the extent of their Ly~$\alpha$ emission, indicating
  the presence of vigorous star formation. The ALMA sources in the two
  giant LABs in our sample are offset from the X-ray AGNs but are
  likely to be physically associated with the LABs. The SFRs of the
  LABs are comparatively modest ($\approx$~150--410~$M_{\odot}$) and
  are consistent with those expected for co-eval massive star-forming
  galaxies. Furthermore, the giant LABs do not host more luminous star
  formation than the smaller LABs, despite being an order of magnitude
  more luminous in Ly~$\alpha$ emission.

\item On the basis of the star formation and Ly~$\alpha$ luminosity of
  the LABs we conclude that star formation can power the extended
  Ly~$\alpha$ emission (through photoionisation) for all of the LABs
  explored here, although the escape fraction of 200--912~\AA\ photons
  will need to be relatively high for the two giant LABs (LAB~2 and
  LAB~3). However, since the ALMA sources in the giant LABs are offset
  from the centre of the Ly~$\alpha$ emission and the X-ray AGN it is
  also possible that multiple systems photoionise the Ly~$\alpha$
  emission and deeper ALMA observations may reveal fainter
  star-forming galaxy components.

\end{itemize}

Overall, our study has provided a mixed message on the mean SFRs of
X-ray AGNs in a protocluster environment. From our original ALMA
sample, which explored a range of regions across the protocluster,
there was no strong evidence for a significant enhancement in mean SFR
for the protocluster AGNs over that found for AGNs in the
field. However, when including ALMA data for AGNs in the protocluster
core from Umehata et~al. (2015), evidence was found for elevated mean
SFRs over the field, although the mean SFR was dominated by one
exceptional protocluster AGN with a SFR $\simgt$~3 times higher than
the other AGNs. Our results therefore provide evidence that star
formation is enhanced for the AGNs in the central region of the
protocluster but is consistent with field AGNs outside of this central
region. To more comprehensively measure how much the protocluster
environment effects star formation would require a complete census of
star formation across the whole of the protocluster for both AGNs {\it
  and} galaxies. Mapping the protocluster with ALMA would achieve this
aim and, when combined with the deep {\it Chandra} observations, would
also allow for detailed BH--galaxy growth comparisons across the full
protocluster environment.

\section*{acknowledgments}

We thank the referee for a constructive and useful review of the
paper. We gratefully acknowledge support from the Leverhulme Trust
(DMA), the Science and Technology Facilities Council (DMA; CMH; IRS;
DJR; AMS; ST/L00075X/1), the ERC Advanced Investigator grant DUSTYGAL
321334 (JMS; IRS), a Royal Society Wolfson Merit Award (IRS), the
Collaborative Research Council 956, sub-project A1, funded by the
Deutsche Forschungsgemeinschaft (AK), KAKENHI (YM; 20647268), the
Durham Doctoral Scholarship (FS), and the Grant-in-Aid for JSPS
Fellows (HU; 26.11481). This paper makes use of the following ALMA
data: ADS/JAO.ALMA\#2011.0.00725.S\*. ALMA is a partnership of ESO
(representing its member states), NSF (USA) and NINS (Japan), together
with NRC (Canada) and NSC and ASIAA (Taiwan), in cooperation with the
Republic of Chile. The Joint ALMA Observatory is operated by ESO,
AUI/NRAO and NAOJ.


{}

\end{document}